# An Exploration of the Physics of Spherically Symmetric Dynamic Horizons

James Lindesay

*Computational Physics Laboratory, Howard University*
*Washington, D.C.*

**Abstract.** Geometries with horizons offer insights into relationships between general relativity and quantum physics. For static spherically symmetric space-times, the event horizon is coincident with a coordinate anomaly that introduces complications in descriptions of near horizon physics. Naïve introduction of dynamics using coordinates with anomalous behavior coincident with the horizon also introduces invariant singular physical content at that horizon. However, the introduction of a temporal coordinate that is non-orthogonal to spatial coordinates near the horizon, but asymptotically orthogonal, provides a dynamic description without singular physical content at the horizon itself. Penrose diagrams will be presented exhibiting temporal dependencies for accreting and evaporating black holes, and near horizon light-like trajectories will be examined. In addition, the quantum mechanics of simple quantum fields will be explored. Finally, a two-fluid cosmology will be suggested to describe dynamic coherent aspects of the universe as a whole.

**Keywords:** Dynamic black holes, dark energy, Penrose diagrams.
**PACS:** 04.62.+v 04.70.-s 95.30.Sf 98.80.Jk

## INTRODUCTION

There is considerable interest in the interplay of quantum mechanics with the space-time arenas used to describe large-scale gravitating phenomena. Horizons, being defined by light-like surfaces, can serve as conceptual laboratories for the examination of the causal structures of space-times, as well as any coherent behaviors of their physical content. However, since horizons can only be defined globally in some circumstances, one does not expect local anomalous physical behavior to be associated with horizons in any inherent manner. If one's intuition is based upon the behavior of static horizons, the temporal dynamics of physical systems near such horizons is often anomalous due to the singular behavior of that temporal coordinate. Since one expects phenomena such as black holes to be able to have dynamic masses, care should be taken on the choice of the temporal parameter used to describe that dynamics in order to insure that unphysical behaviors are not introduced due only to incorrect parameterization of the dynamics.

This paper will examine the dynamics of systems with spherically symmetric horizons using locally a non-orthogonal temporal parameter that is non-singular near the horizon, and corresponds with the Minkowski time parameter for an asymptotic observer. In order to develop a framework for this exploration, a brief introduction will be given on representations of global space-times.

# Space-Time Diagrams

A standard space-time diagram represents time as the vertical coordinate axis, space as the horizontal coordinate axis, and is scaled such that light-like curves have a slope of one when plotted on the diagram. Although such diagrams are intuitively straightforward and useful, they cannot represent the large-scale structure of the space-time on a single page. Penrose diagrams have been developed for this purpose[1]. The coordinates of a Penrose diagram are chosen to satisfy two conditions:

- Penrose diagrams preserve the flat space-time slope of light-like curves. Outgoing photon crosses equal r and ct coordinates,
- Penrose diagrams map infinite space-time coordinates onto a finite page.

As an example, consider flat Minkowski space-time. The metric form describing invariant distances in the space-time is given in spherical polar coordinates by

$$ds^2 = -c^2 dt^2 + dr^2 + r^2 d\theta^2 + r^2 \sin^2 \theta d\varphi^2 \qquad (1)$$

There is some arbitrariness on how one chooses to map the coordinates onto a finite page. In what follows, hyperbolic tangents will be used for this coordinate map. To represent Minkowski space, the horizontal coordinate will be given by

$$Y_h = \frac{-Tanh(ct-r)+Tanh(ct+r)}{\sqrt{2}} \qquad (2)$$

while the vertical coordinate is given by

$$Y_v = \frac{Tanh(ct-r)+Tanh(ct+r)}{\sqrt{2}} \qquad (3)$$

The representations of Minkowski space-time are shown in FIGURE 1:

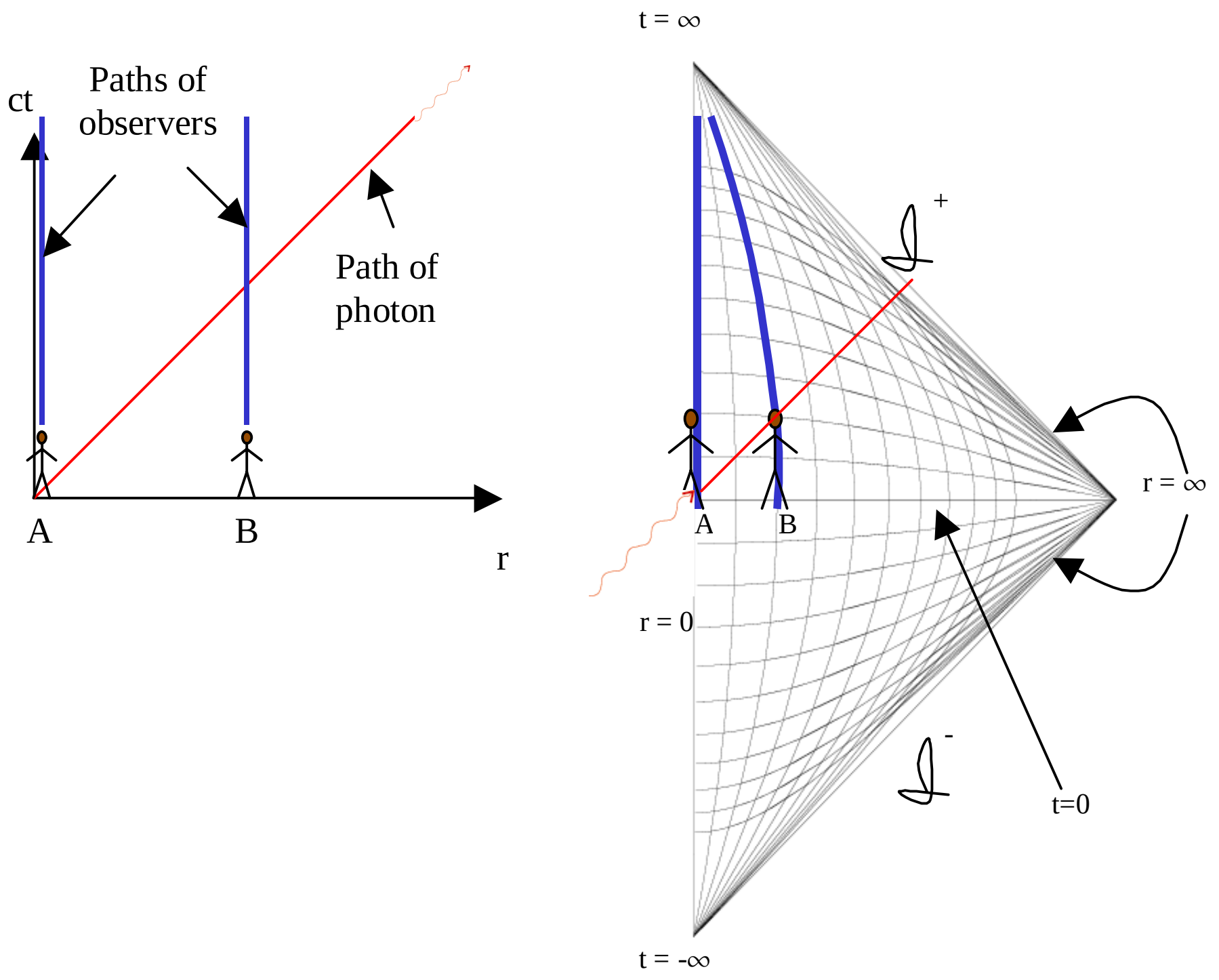

**FIGURE 1** Representations of Minkowski space-time. The Penrose diagram is on the right.

The Penrose diagram is seen to represent asymptotic regions in the space-time as labeled. The path of the photon is also seen to maintain a slope of unity in the diagram. This is quite convenient for examining the large-scale causal structure of the space-time, since causally connected regions are separated by light-like surfaces.

## A Static Black Hole

Another example space-time is given by the Schwarzschild black hole. For conciseness of expression, define the Schwarzschild radius as the constant $R_S \equiv \dfrac{2MG}{c^2}$. The space-time metric form is then given by

$$ds^2 = -\left(1 - \frac{R_S}{r}\right)c^2 dt^2 + \frac{dr^2}{\left(1 - \frac{R_S}{r}\right)} + r^2 d\theta^2 + r^2 \sin^2\theta d\varphi^2 \qquad (4)$$

This is the general form of a static spherically symmetric space-time, describing motions external to gravitating spherical stars, planets, etc. The coordinates are seen to be asymptotically flat, i.e., the space-time looks like Minkowski space-time as $r \rightarrow \infty$. If the vacuum solution holds down to r=0, then the geometry has a coordinate singularity at $r=R_S$. This coordinate singularity at the Schwarzschild radius is a light-like surface that also corresponds to the horizon of the black hole. Since the horizon is a light-like surface, any observer static in the coordinate $r=R_S$ must experience singular proper acceleration (along with the forces needed to maintain this acceleration) at this coordinate. For general observers, the physical tidal forces experienced upon crossing the horizon depend on the space-time curvature components, which all scale like $(1/R_S)^2$ and are therefore finite. Therefore this is not a physical singularity. However, the singularity at $r$=0 is physical.

The coordinates for the Penrose diagram can be developed in terms of a transformation to conformal coordinates that preserve the slope of light curves. The horizontal coordinate that will be used for the diagram is given by

$$Y_h = \frac{-Tanh\left(\frac{ct-\left(r+R_S\log\left(\frac{r}{R_S}-1\right)\right)}{R_S}\right)+Tanh\left(\frac{ct+\left(r+R_S\log\left(\frac{r}{R_S}-1\right)\right)}{R_S}\right)}{\sqrt{2}} \quad (5)$$

and the vertical coordinate is given by

$$Y_v = \frac{Tanh\left(\frac{ct-\left(r+R_S\log\left(\frac{r}{R_S}-1\right)\right)}{R_S}\right)+Tanh\left(\frac{ct+\left(r+R_S\log\left(\frac{r}{R_S}-1\right)\right)}{R_S}\right)}{\sqrt{2}}. \quad (6)$$

The Penrose diagram is demonstrated in FIGURE 2:

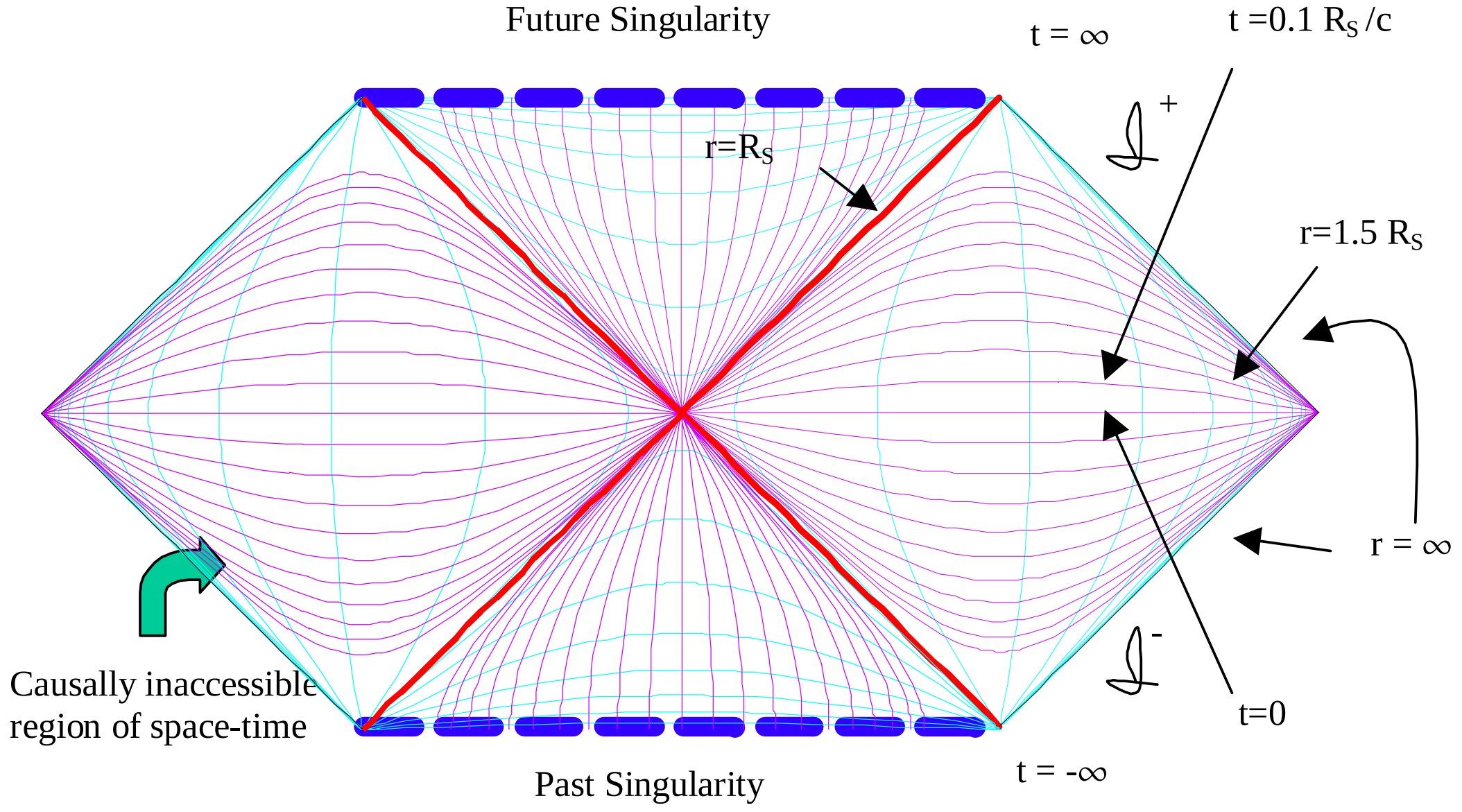

**FIGURE 2.** Penrose diagram for a Schwarzschild black hole.

The region on the right is the causal patch accessible to an observer external to the black hole, whereas the region on the right is included to maximally complete the causal structure of the geometry. The structure of the space at $r=\infty$ mimics the Minkowski Penrose diagram from before. For the static Schwarzschild geometry, the singularity $r=0$ is the space-like horizontal dashed line at the top of the diagram. The light-like horizon $r=R_S$ is also a $t=\infty$ surface, meaning that no finite fixed time curves penetrate this surface on the diagram. In the next sections, geometries with temporally dynamic horizons will be explored.

# TEMPORALLY DEPENDENT BLACK HOLES

## Dependence of Mass on Schwarzschild Time

One expects that if there can be accretion and evaporation of a black hole, then the mass of a black hole, and all relevant scales, must assume a temporal dependency. There are various choices one could make to introduce this dynamic behavior. If one introduces a naïve dependence of the mass of a Schwarzschild-like black hole upon the Schwarzschild time variable, a direct calculation of the physical characteristics of the resulting space-time indicates a physical singularity at the now dynamic Schwarzschild radius. One can define a radial mass scale $R_M$ that describes the geometry of such a black hole with a time dependent mass. The Ricci scalar can then be directly calculated:

$$R_M(ct_S) \equiv \frac{2GM(ct_S)}{c^2} \quad ,$$

$$ds^2 = -\left(1 - \frac{R_M(ct_S)}{r}\right) c^2 dt_S{}^2 + \frac{dr^2}{\left(1 - \frac{R_M(ct_S)}{r}\right)} + r^2 d\theta^2 + r^2 \sin^2\theta d\varphi^2 \quad , \tag{7}$$

$$\Re = -\frac{r\left(2\dot{R}_M^2 + (r - R_M)\ddot{R}_M\right)}{(r - R_M)^3} = \frac{8\pi G}{c^4} g^{\mu\nu} T_{\mu\nu} \quad .$$

In this expression, dots represent derivatives with respect to the dynamical variable $ct_S$. Since the Ricci scalar is directly related to invariant physical content (the trace of the energy-momentum tensor), this means that the singular behavior in this scalar at $r=R_M$ reflects singular physical behavior for all observers on or passing this surface.

## Use of Non-Orthogonal Coordinates

One is able to construct a dynamic black hole geometry using spatio-temporal coordinates that are non-orthogonal "near" the black hole.

$$ds^2 = -c^2dt^2 + \left[ dr + \sqrt{\frac{R_M(ct)}{r}} cdt \right]^2 + r^2 d\theta^2 + r^2 \sin^2 \theta d\varphi^2 . \qquad (8)$$

The asymptotic form of this metric is that of Minkowski space-time. If the radial mass scale $R_M$ was a constant equal to the Schwarzschild radius, one can directly diagonalize the form yielding a Schwarzschild metric[2]. Again, the Ricci scale can be directly calculated for this geometry, where it takes the form[3,4,5,6]

$$\Re = -\frac{3\dot{R}_M}{2r^2}\sqrt{\frac{r}{R_M}} = \frac{8\pi G}{c^4} g^{\mu\nu} T_{\mu\nu} , \qquad (9)$$

and now the dot represent a derivative with respect to the dynamical variable *ct*. Note that for this geometry the physical content invariant is everywhere non-singular away from *r=0*. The spatio-temporal mixing in the coordinates describing regions near the horizon has eliminated physically singular behavior away from the dynamic mass at the center of the geometry. The Ricci scalar is seen to vanish identically when $\dot{R}_M = 0$.

Because the mass scale is dynamic, one expects the horizon in this space-time to likewise be dynamic. A black hole's horizon is defined by an outgoing radial light-like (null) geodesic. The general form for radial null surfaces in this geometry is given by

$$\begin{gathered} ds_\gamma{}^2 = 0 = -c^2 dt^2 + \left[ dr_\gamma + \sqrt{\frac{R_M(ct)}{r_\gamma}} cdt \right]^2 + r_\gamma{}^2 d\theta_\gamma{}^2 + r_\gamma{}^2 \sin^2 \theta_\gamma d\varphi_\gamma{}^2 , \\ d\theta_\gamma = 0 = d\varphi_\gamma \end{gathered} \qquad (10)$$

This means that radial outgoing/ingoing photons are expected to satisfy

$$\frac{dr_\gamma}{dct} = -\sqrt{\frac{R_M}{r_\gamma}} \pm 1 . \qquad (11)$$

The horizon consists of the outermost set of null geodesics that cannot reach future light-like infinity $\Im^+$. In particular, the horizon satisfies

$$\frac{dR_H}{dct} = -\sqrt{\frac{R_M}{R_H}} + 1 \quad , \quad R_H = \frac{R_M}{\left(1 - \dot{R}_H\right)^2} \begin{cases} > R_M & accretion \\ < R_M & evaporation \end{cases} , \qquad (12)$$

and is seen to lie external to the radial mass scale during accretion, and internal to this scale during evaporation. Several characteristics of the horizon are particularly noteworthy:

- Unlike static Schwarzschild geometry, $R_H \neq R_M$.
- Radially traveling photons located at $r=R_M$ are momentarily stationary, while a dynamic horizon is not!
- Radially outgoing photons between $R_H < r_\gamma < R_M$ will escape the singularity!

## Construction of the Penrose Diagram

Next, in order to construct the Penrose diagram for this space-time, one needs to develop conformal coordinates such that light moves on 45° lines on the diagram. The coordinates should be orthogonal in temporal and radial coordinates. Any transformation to conformal coordinates must insure integrability on mixed partial derivatives with respect to the original coordinates. The Penrose diagram is then constructed by mapping these conformal coordinates onto a finite space[6].

The metric will take the following form when expressed in terms of the conformal coordinates:

$$ds^2 = \frac{-(dct_*)^2 + dr_*^2}{\tilde{W}_+ \tilde{W}_-} + r^2\left(d\vartheta^2 + \sin^2\vartheta\, d\varphi^2\right). \tag{13}$$

Fortunately, for the case that the rate $\dot{R}_M$ is constant (i.e. $\ddot{R}_M = 0$), the form of the conformal coordinates and scale factors can be integrated. The conformal coordinates satisfy

$$r_* \pm ct_* = r \exp\left[\int_r^{R_M(ct)} \frac{\left(1 \pm \sqrt{\zeta'}\right) d\zeta'}{\zeta'\left(1 \pm \sqrt{\zeta'}\right) \pm \dot{R}_M}\right]. \tag{14}$$

The conformal scale factors can be expressed in terms of a reduced parameter $\zeta(ct,r)$, which takes on a constant value $\zeta_H$ for the horizon:

$$\begin{aligned} \zeta &\equiv \frac{R_M(ct)}{r}, \\ \zeta_H\left(1 - \sqrt{\zeta_H}\right) - \dot{R}_M &= 0 \text{ for the horizon.} \end{aligned} \tag{15}$$

The conformal scale factors in Eq. (13) satisfy

$$\tilde{W}_\pm = \frac{\pm \dot{R}_M}{\zeta\left(1 \pm \sqrt{\zeta}\right) \pm \dot{R}_M} \exp\left[\int^{\zeta} \frac{\left(1 \pm \sqrt{\zeta'}\right) d\zeta'}{\zeta'\left(1 \pm \sqrt{\zeta'}\right) \pm \dot{R}_M}\right]. \tag{16}$$

The Penrose diagram for this geometry is demonstrated in FIGURE 3.

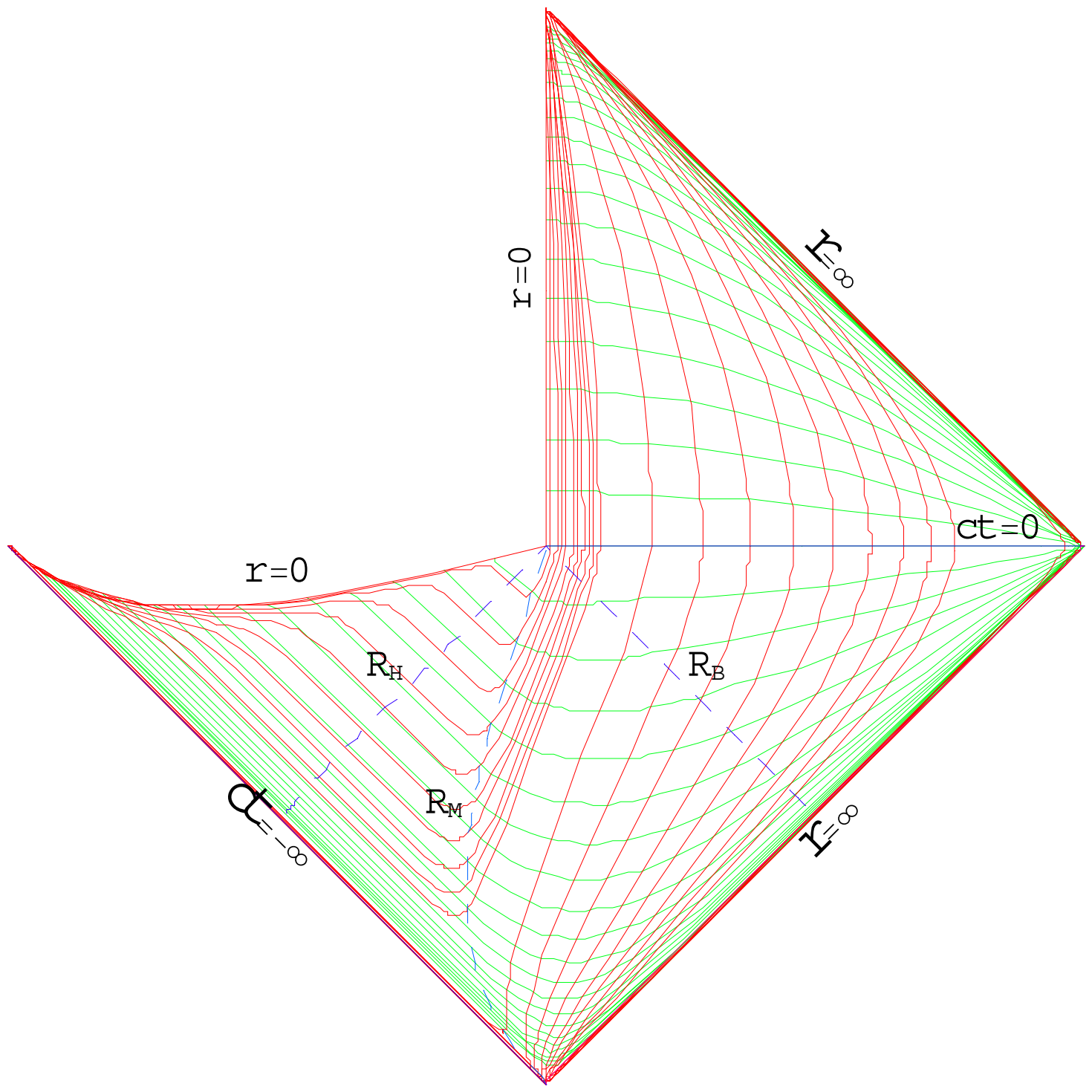

**FIGURE 3.** Penrose diagram for evaporating black hole.

The constant *ct* curves run more or less horizontally in the right-hand portion of the diagram, while the constant *r* curves run more or less vertically in that region. All constant *ct* curves originate on the curve $r$=0 and terminate in the right corner of the diagram $r=\infty$. The constant *r* curves originate in the left corner of the diagram $ct=-\infty$ and terminate in the top corner $ct=\infty$. For illustrative purposes, the spacing of the constant *r* curves are originally in units of tenths until $r$=1. The space-time above the horizontal curve $ct$=0 is Minkowski space-time. The dashed line labeled $R_H$ in the lower left region of the diagram represents the dynamic horizon, while the dashed line labeled $R_B$ in the lower right region of the diagram represent the outermost ingoing light-light surface reaching the singularity prior to the completion of evaporation. The dashed curve labeled $R_M$ origination at the bottom corner of the diagram and terminating at the point of termination of the singularity $(ct=0, r=0)$ is the radial mass scale of the black hole.

It might be somewhat surprising that a majority of the Penrose diagram of a black hole that evaporates at a constant rate (and therefore has an unbounded area in the distant past) represents space-time external to the horizon. The radial scales of the regions external to the black hole are more unbounded than the horizon scale of that black hole.

## Photon Trajectories

It is informative to examine the trajectories of radially moving photons in this dynamic space-time[7]. These photons satisfy Eq. (11), and if the transformation to conformal coordinates is accurate, their trajectories should be represented as lines of slope unity on the Penrose diagram. These trajectories are represented in FIGURE 4.

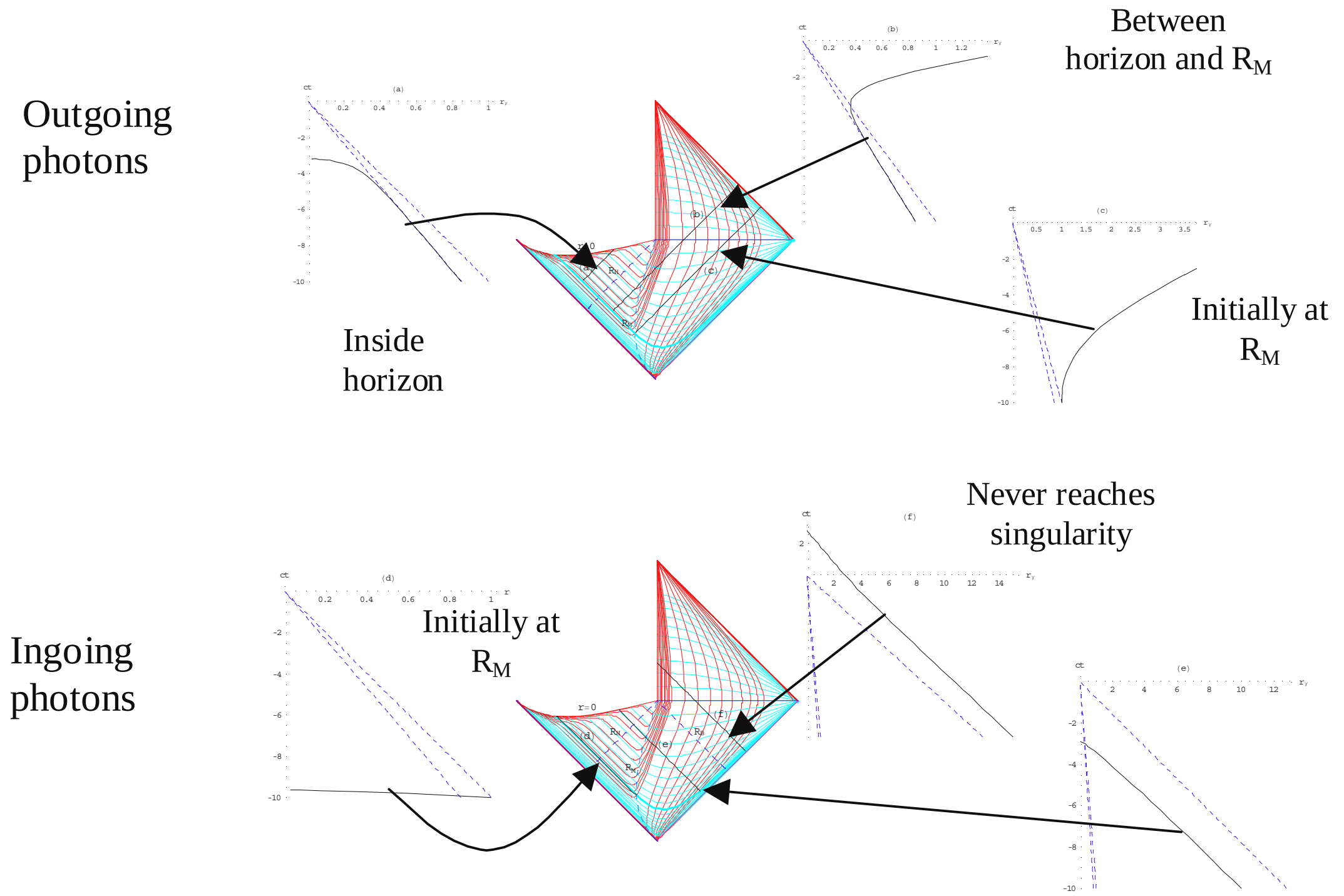

**FIGURE 4.** Radially moving photons near the horizon of an evaporating black hole.

All photons in the diagrams originate at the same time, but differing initial positions and/or directions. As seen in the diagram, because of the near-field gravitational attraction, outgoing photons within the radial mass scale initially approach the singularity. However, photons between the horizon and the radial mass scale eventually reverse their direction as the attraction weakens sufficiently. The outgoing photon that originates at the radial mass scale is initially stationary, while the radial mass scale moves away from it, allowing it to begin moving outwards in the radial parameter $r$. The ingoing photon originating from this point reaches the singularity in a very short temporal interval. One also sees a class of ingoing photons that will reach $r=0$ only after the singularity has vanished.

## QUANTUM MECHANICS IN AN EVAPORATING BLACK HOLE GEOMETRY

The analytic properties of the space-time of the previous evaporating black hole are convenient for examining the behavior of quantum systems. There are well established methods for doing quantum physics in curved space-times[1]. As an example, consider a non-interacting (but gravitating) scalar field. For the geometries being discussed in this paper, it is convenient to decompose the field using[3]

$$\chi_{\ell m}(ct, r, \vartheta, \varphi) \equiv \frac{\psi_\ell(ct, r)}{r} Y_{\ell m}(\vartheta, \varphi) . \quad (17)$$

The Euler-Lagrange equations of motion for a gravitating scalar field are given by

$$-\frac{\partial^2\psi_\ell}{(\partial ct)^2}+\frac{\partial}{\partial ct}\left[\sqrt{\frac{R_M}{r}}\left(\frac{\partial\psi_\ell}{\partial r}\right)\right]+\frac{\partial}{\partial r}\left[\sqrt{\frac{R_M}{r}}\left(\frac{\partial\psi_\ell}{\partial ct}\right)\right]+\frac{\partial}{\partial r}\left[\left(1-\frac{R_M}{r}\right)\left(\frac{\partial\psi_\ell}{\partial r}\right)\right]+$$
$$-\left[\frac{m^2+\ell(\ell+1)+\frac{R_M}{r}}{r^2}+\frac{1}{r}\frac{\partial}{\partial ct}\sqrt{\frac{R_M}{r}}\right]\psi_\ell=0.$$

(18)

It is convenient to define the dimensionless reduced parameter $\zeta\equiv\frac{R_M(ct)}{r}$. This equation can then be rewritten in terms of the single parameter $\zeta$:

$$\left(-\zeta+\zeta^{3/2}+\dot{R}_M\right)\left(\zeta+\zeta^{3/2}+\dot{R}_M\right)\frac{d^2\psi_\ell}{d\zeta^2}+2\zeta^2\left(\zeta(-2+3\zeta)+3\sqrt{\zeta}\dot{R}_M\right)\frac{d\psi_\ell}{d\zeta}+$$
$$\left(2\ell(\ell+1)\zeta^2+2\zeta^3+2m^2R_M^2+\zeta^{3/2}\dot{R}_M\right)\psi_\ell=0.$$

(19)

All numerical work presented will involve $\ell=0$ (s-waves). The energy density can be calculated once the field is known using:

$$T^{00}=\frac{1}{8\pi}\left[|\chi'|^2+|\dot{\chi}|^2+m^2|\chi|^2+\frac{R_M}{r}|\chi'|^2-2\sqrt{\frac{R_M}{r}}\,\Re e\left(\dot{\chi}^*\chi'\right)\right], \qquad (20)$$

where dots represent derivatives with respect to *ct*, and primes are derivatives with respect to *r*. For the present, only massless fields *m=0* will be examined. At the black hole-Minkowski space transition, one expects the quantum field to satisfy appropriate boundary conditions matching the free field solutions in the two spaces. This transition occurs at t=0 in FIGURE 3. The appropriate form of the free massless scalar field in Minkowski space-time is given by

$$\psi_M(ct,r)=\alpha\log\left(\frac{r+ct}{r-ct}\right). \qquad (21)$$

A convenient form to present the numerical behaviors of densities calculated in this geometry is in terms of density plots. The computer is instructed to place a pixel at a point on the Penrose diagram corresponding to the asymptotic observer's coordinate label *(ct, r)* if the density multiplied by a random number between 0 and 1 is larger than the value calculated at a normalization point. Such a plot demonstrates relative measures of the density throughout the space-time. As an example, consider the Ricci scalar for the evaporating black hole, given in Eq. (9). This scalar density, which is directly related to invariant physical content, is plotted in FIGURE 5.

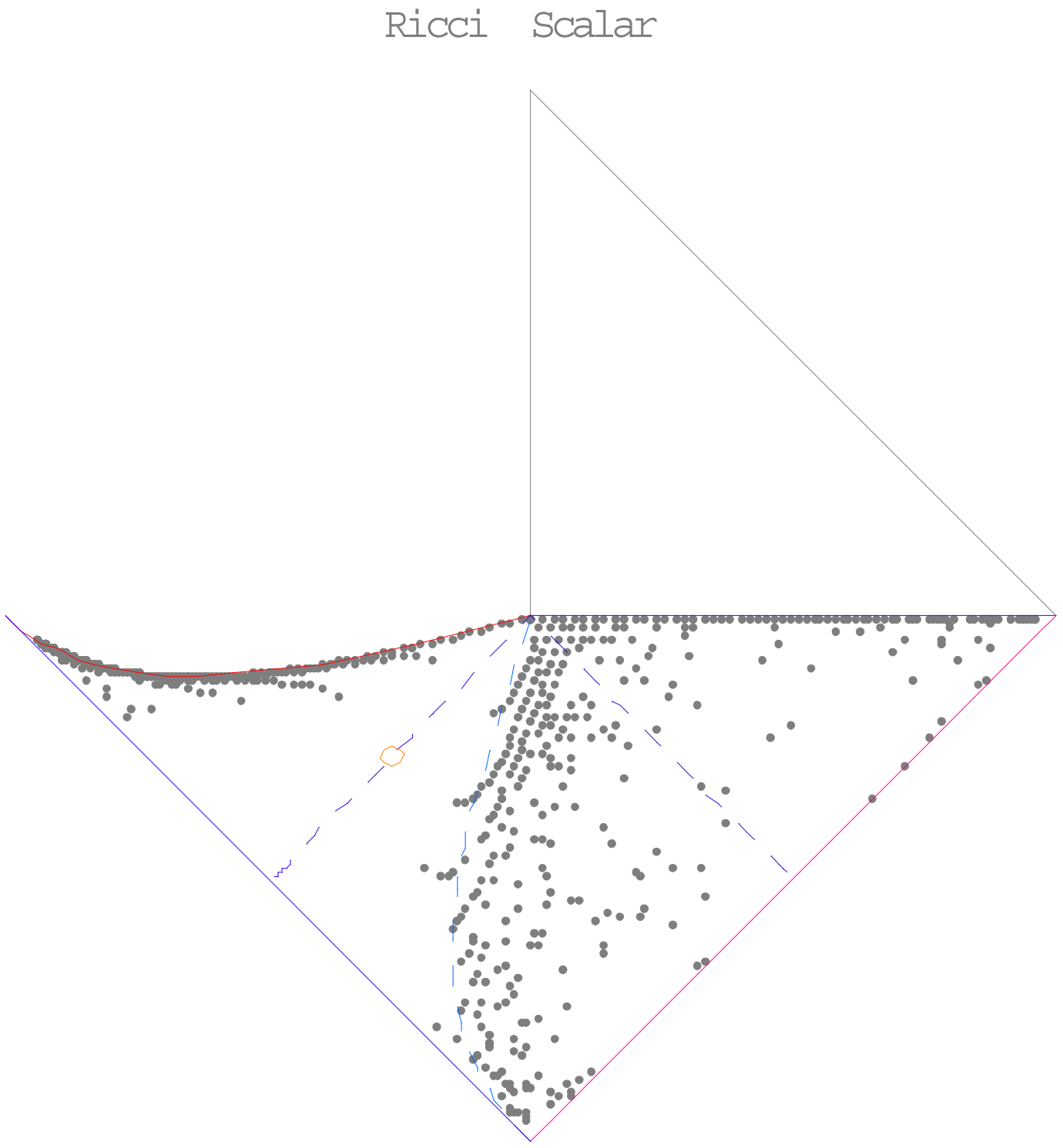

**FIGURE 5** Preliminary density plot of Ricci scalar.

The normalization density for this diagram is indicated by the center of the small circle in the middle of the dashed line in the lower left region of the Penrose diagram representing the horizon of the black hole. Of course, the Ricci curvature scalar vanishes in the upper right region of the diagram representing the Minkowski space-time portion of the geometry.

One can similarly represent the probability densities generated by solutions to the quantum field equations for the s-wave of the massless scalar field satisfying Eq. These densities are exhibited in FIGURE 6:

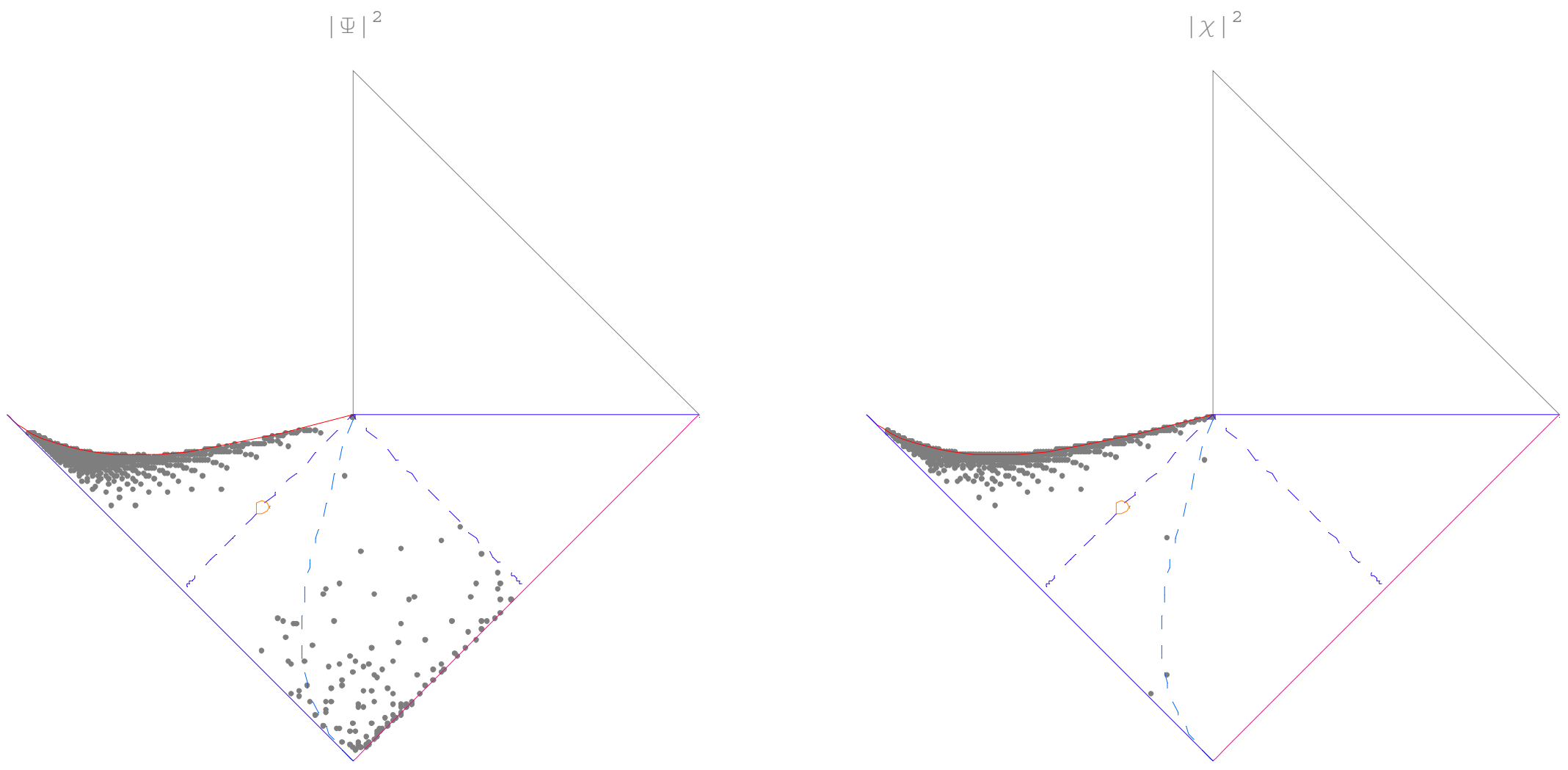

**FIGURE 6** Preliminary probability density plots for massless scalar field in terms of reduced radial wavefunction and radial wave function $\chi = \psi / r$.

Once again, the densities are normalized on the horizon at the point indicated by the small circle.

One ultimately desires a self-consistent quantum field that actually generates the geometry, not just satisfies the dynamical equations on the black hole space-time background. For this to be achieved, a proper quantum field satisfying proper boundary conditions must be developed. One might gain a clue towards the form of the consistent field by examining the invariant physical content generated by the field, and comparing it to the Ricci scalar of the geometry. The invariant physical content (trace of the energy-momentum tensor) for the s-wave massless scalar field satisfying non-vanishing boundary conditions for $\zeta=0$ is shown in FIGURE 7:

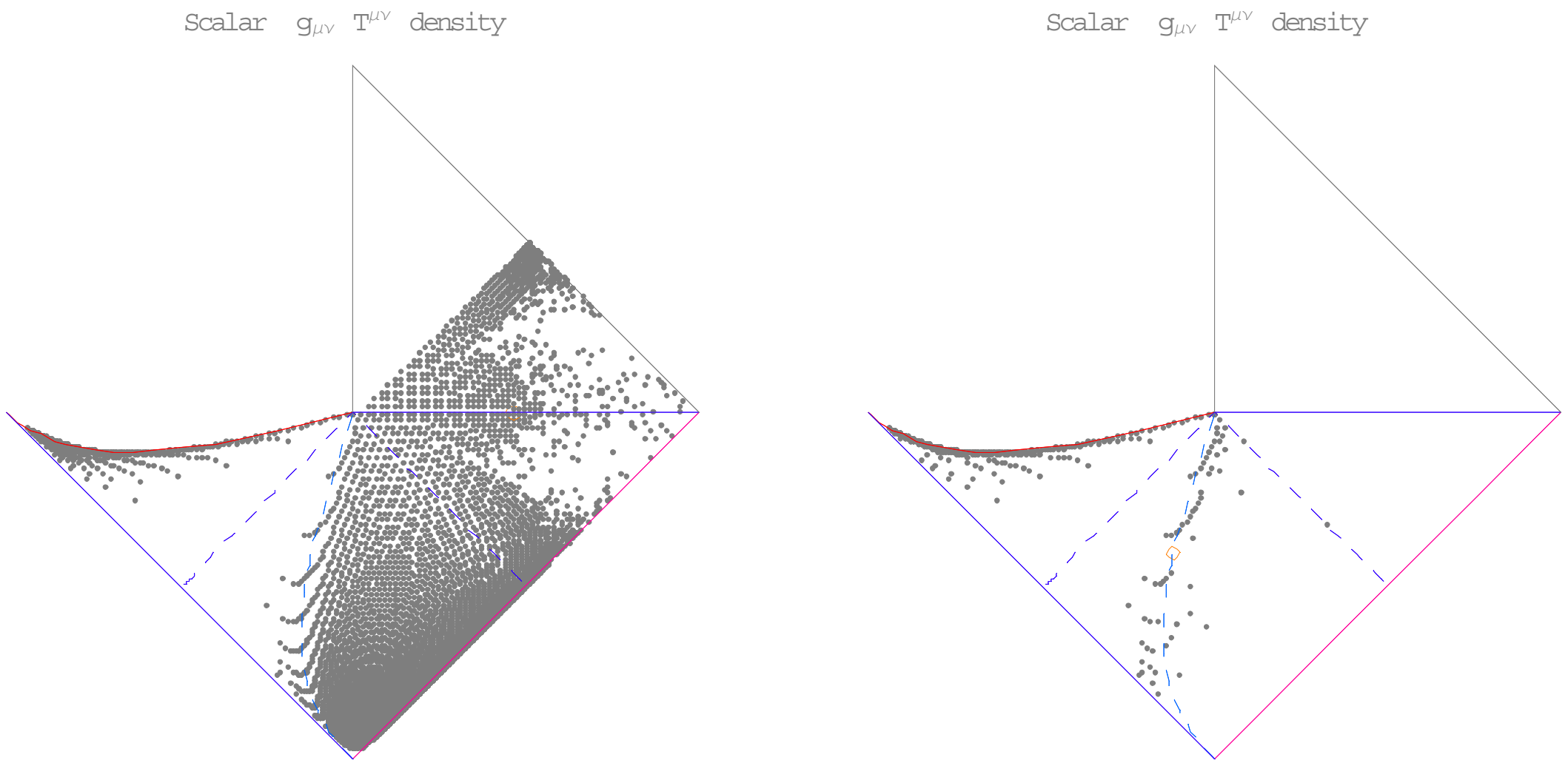

**FIGURE 7** Preliminary density plot of invariant physical content $g_{\mu\nu}T^{\mu\nu}$ of an example scalar field with non-vanishing value at the black hole-flat space interface.

The two diagrams refer to the same physical system, just choosing differing normalization points. The first diagram demonstrates that boundary conditions on the quantum field at *t*=0 are indeed properly satisfied. The second diagram normalized on the radial mass scale indicates that the size of the invariant is large near the horizon and radial mass scale. These behaviors will be explored in future work by the author.

# LIFE CYCLE OF A BLACK HOLE

The black hole examined in the previous two sections evaporates at a steady rate. However, one expects real black holes to vary in rates of accretion and evaporation. This behavior can be approximated by constructing piecewise steady accretions or evaporations, and properly matching the geometries across transition times. To begin, stepwise steady accretion into a black hole will be examined.

## Stepwise Steady Accretions

The conformal coordinates found in Eq. (14) require that the rate of accretion or evaporation $\dot{R}_M$ be constant. However, these coordinates can be utilized if the areas of spheres of radius $r$ are maintained across transitions between periods of constant rate. The consistency of transverse lengths and areas between the geometries is required due to the identity of the transverse part of the space-times $ds_\perp^2 \equiv r^2 \left( d\vartheta^2 + \sin^2 \vartheta d\varphi^2 \right)$ amongst the various patches within the geometry.

As was the case for the final state of the evaporating black hole, the accretion will be assumed to begin at t=0 from initially flat space-time. The example system that was examined followed transitions as indicated in FIGURE 8.

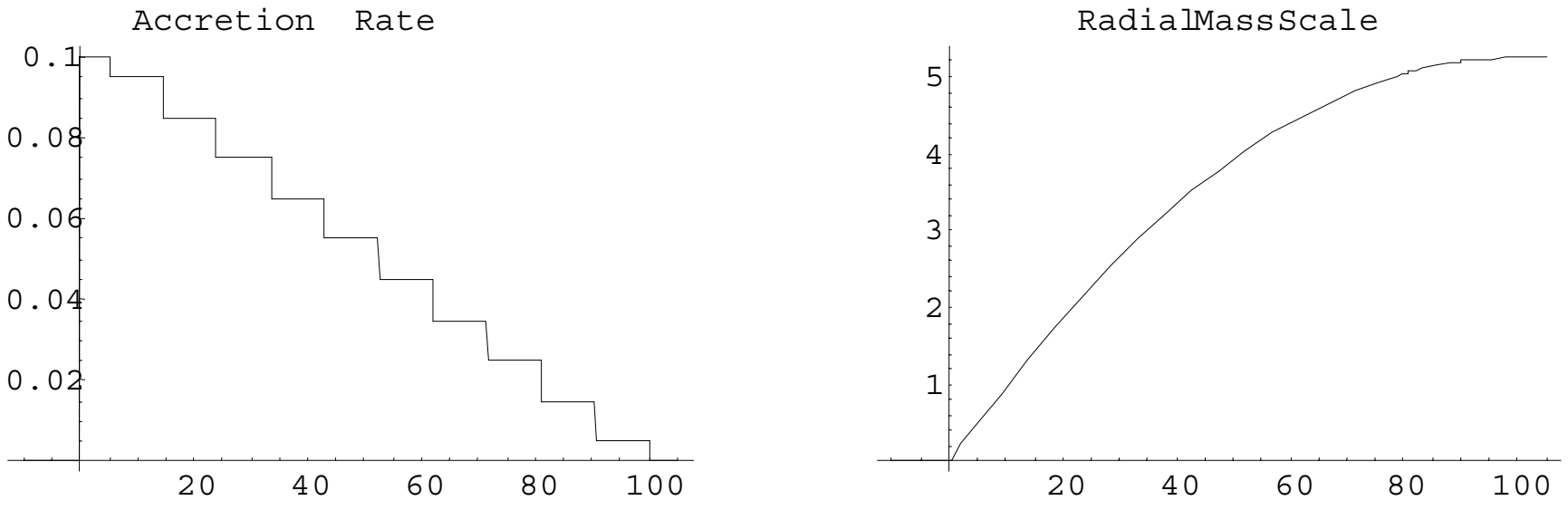

**FIGURE 8** Example rate steps $\dot{R}_M$ and radial mass scale of stepwise steady accretions.

The plot on the left demonstrates an initial transition from a geometry with no black hole, to a black hole that undergoes steps in the accretion rate. The plot on the right indicates the size of the radial mass scale, which is proportional to the mass of the dynamic black hole. The transition that initiates the formation of the physical singularity at *t*=0 will be of particular interest. The resulting Penrose diagram is shown in FIGURE 9.

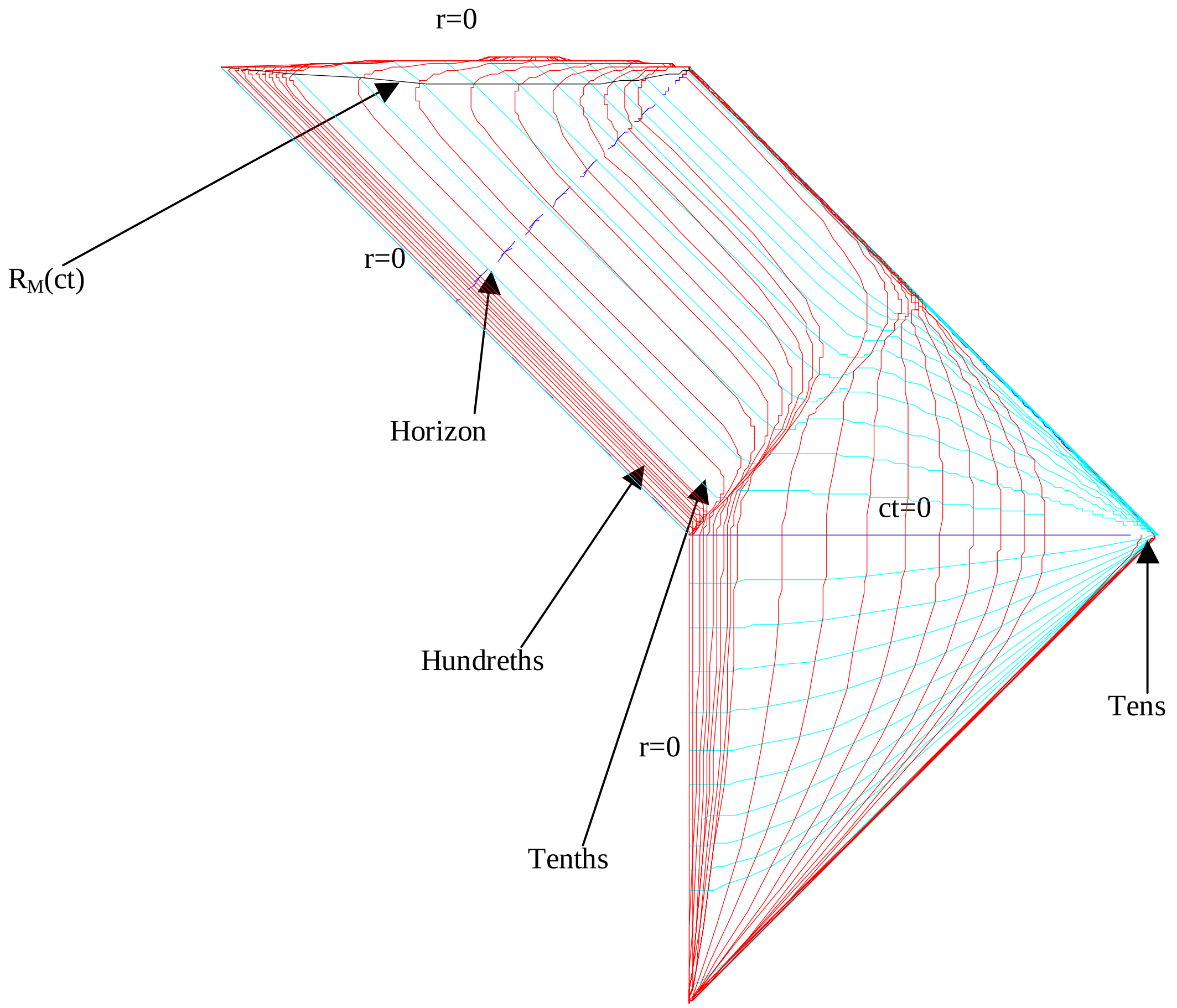

**FIGURE 9** Preliminary Penrose diagram for a black hole with piecewise steady accretion.

Several features of this diagram are of interest. The time slice *t=0* is labeled as the darker horizontal line separating the upper and lower right-hand sections of the diagram. First, as expected, the coordinate *r=0* which is the time-like left border of the lower right Minkowski space region becomes the space-like upper bound of the upper left region of the diagram after the singularity forms. It is quite interesting that this coordinate undergoes an extended light-like transition on the diagram during the formation of the singularity. The horizon is the dashed light-like curve in the upper left region of the diagram. It is clearly seen to be dynamic, since constant time and constant radial coordinate *r* curves cross the horizon on the diagram. Also, as required by Eq. (12), the radial mass scale $R_M(ct)$ is a space-like curve between *r=0* and the horizon.

To check the correctness of the conformal coordinates and the numerical solutions, radial light-like trajectories will be plotted. Numerical solutions to Eq. (11) for an outgoing and ingoing photon are plotted on space-time and Penrose diagrams in FIGURE 10:

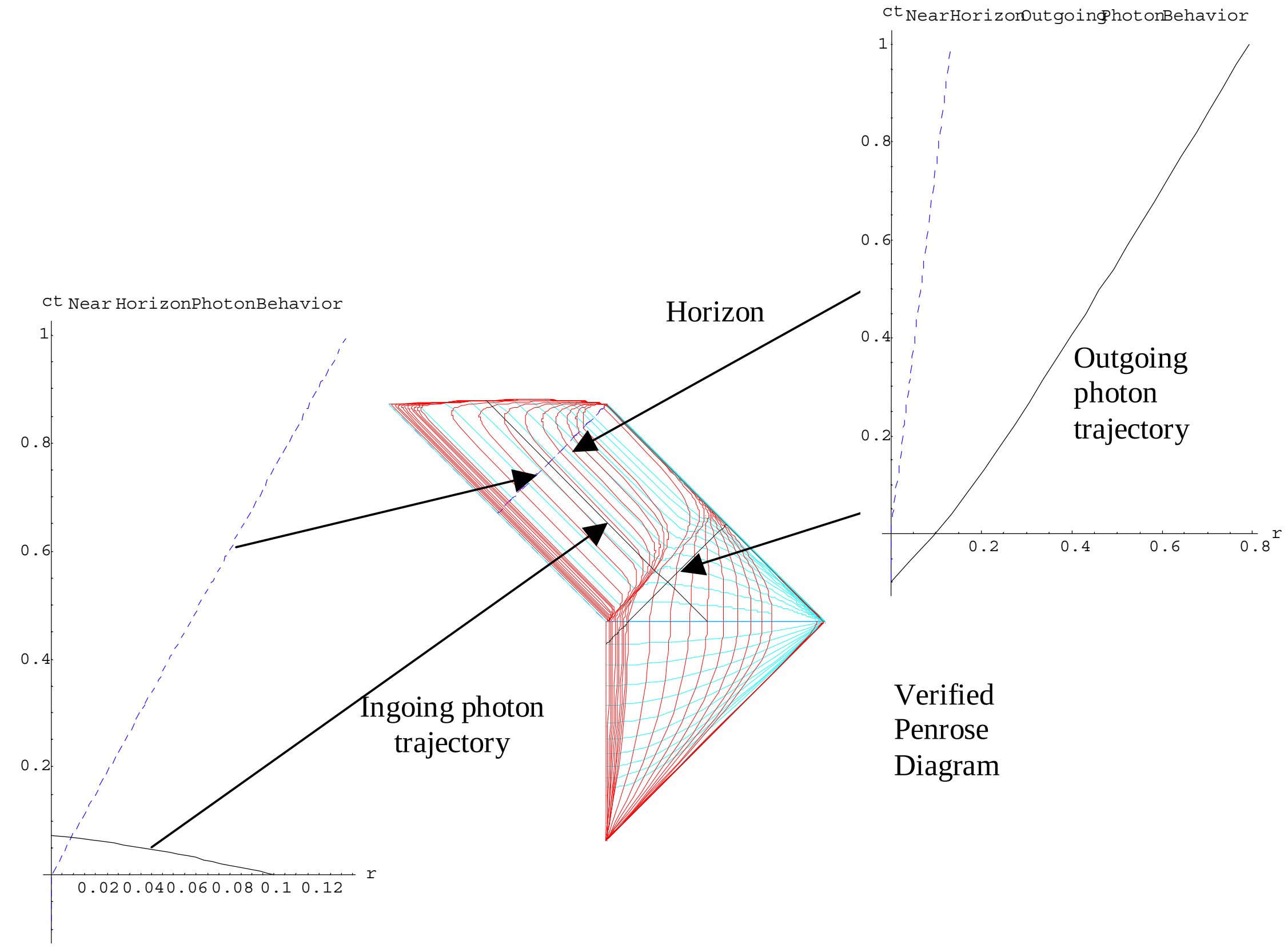

**FIGURE 10** Example photon trajectories in accreting black hole.

# Accretion, Then Evaporation

Of considerable interest is the Penrose diagram for the complete life cycle of a black hole. Consider the formation of a black hole geometry from an initially flat Minkowski space-time. The mass of the black hole undergoes periods of accretions at fixed rates, and then later undergoes periods of evaporations at fixed rates. The mass of the black hole eventually vanishes, and the geometry again evolves into flat Minkowski space-time. The conformal coordinates that have been developed have the advantage of themselves having asymptotic ($r \rightarrow \infty$) flat space-time forms. This means that the initial and final Minkowski space-times should asymptotically match with the behaviors of the accreting and evaporating black hole. A preliminary form for a black hole whose mass dynamics is symmetric about $t=0$ is calculated in FIGURE 11.

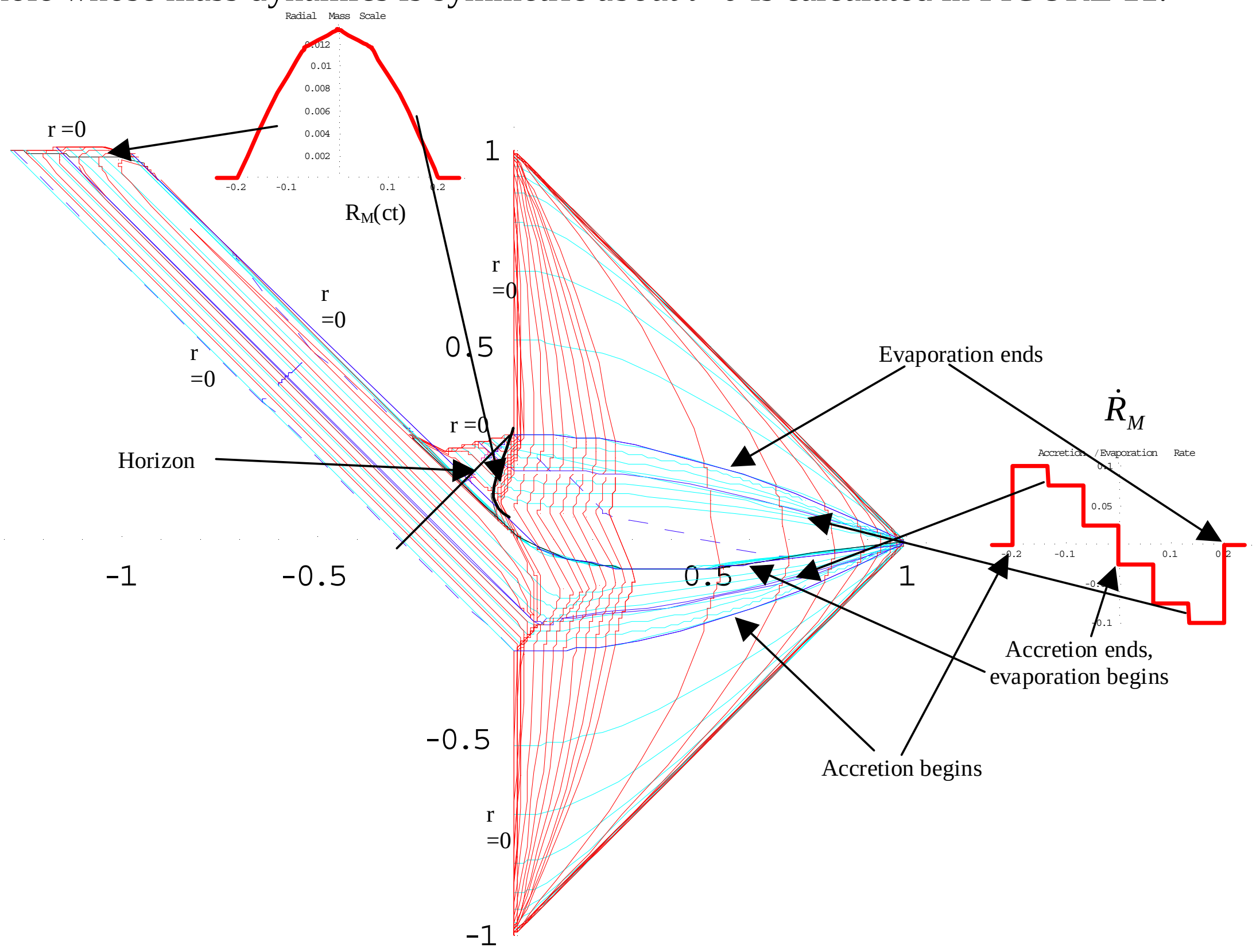

**FIGURE 11** Preliminary Penrose diagram for the complete life cycle of a black hole.

The accretion begins at the lowest dark (blue) fixed time curve indicated on the diagram. The transition from the Minkowski space-time beneath this curve and the accreting black hole above this curve is seen to cleanly preserve transverse lengths and areas through the continuity of the fixed $r$ curves during this transition, as well as the successive transitions indicated by the dark fixed time curves. The boundary conditions are chosen to preserve the coordinate representation of the radial mass scale $R_M(ct)$, which is a physical parameter defining the black hole. The curve $r=0$ at the center of the geometry is initially a time-like trajectory (as expected) in the flat space-time, undergoes a light-like transition as the singularity forms when accretion begins at a negative time to become a space-like trajectory, remains space-like during the black hole epoch of the geometry (although it rapidly changes its geometric representation as accretion becomes evaporation at $t=0$), then once again becomes time-like after the black hole completely evaporates away, leaving flat Minkowski space-time. The radial mass scale $R_M$ is represented by the black curves just beneath $r=0$ during accretion and reflecting $r=0$ across the horizon during evaporation, indicated by arrows from its temporal plot in the figure. As expected from Eq. (12) and previous work,[3,4] the radial mass scale is inside of the horizon during accretion, and outside of the horizon during evaporation. The horizon is seen to be globally defined in terms of the geometry of the late evaporation stage of the black hole, irrespective of the history of the accretions. Fixed radial coordinate curves are seen to cross the horizon and radial mass scale either twice ($r<R_{Hmax}$, $R_{Mmax}$) or not at all, corresponding to growth then shrinkage of these scales during the life cycle of the black hole. However, all fixed time curves are seen to cross these scales only once throughout the history of the black hole. The temporal and radial coordinate mappings cross at unique points as expected for a true coordinate representation. This has certainly not been the case when the author has made computational or representational mistakes in the construction of such diagrams, which supports the viability of the diagrams presented.

## COSMOLOGICAL COHERENCE

The discussion will now shift from the examination of spherically symmetric dynamic black holes to an examination of the cosmology as a whole. A geometry is here referred to as *coherent* if its dynamics maintains distant (supra-luminal) correlations in regions that have not been luminally connected. There is some observational evidence for geometric correlations beyond causally connected regions in cosmology. The observed dark energy distribution indicates a late stage coherently accelerating expansion of the universe. In addition, dark energy introduces a scale in what is an otherwise scale invariant cosmology. It is popular to model this phenomenon using a cosmological constant[8]. In the standard model of cosmology, the cosmological constant must be just that (a constant) in order preserve conservation principles (such as local conservation of energy-momentum). However, some models assert an inflationary initial stage of the universe governed by a differing de Sitter scale than that suggested by the present dark energy density. Of course, any inflation that ends can't really be described by a cosmological constant. It would therefore be convenient to develop a coherent dynamic scale that can be incorporated consistently in a cosmology that incorporates the known big-bang phenomenology. Such a form is exhibited later in this chapter.

### Fluid Cosmology

To provide hints into the general properties of geometries that satisfy the observed phenomena, the problem will be examined in terms of the physical parameters involved. Consider a general cosmology driven by an ideal fluid. The energy-momentum tensor of such a fluid (in a given geometry) is given by the form

$$
\begin{aligned}
T_{\mu\nu} &= P\, g_{\mu\nu} + (\rho + P) u_\mu u_\nu , \\
u_\mu\, g^{\mu\nu} u_\nu &= -1.
\end{aligned}
\tag{22}
$$

The fluid flow will be assumed to be isotropic, implying that $u_\vartheta = 0 = u_\varphi$. One can then turn the problem around and determine the physical fluid parameters associated with any given geometry:

$$
\begin{aligned}
P &= T_\vartheta^\vartheta = T_\varphi^\varphi = -\frac{c^4}{8\pi G_N} G_\vartheta^\vartheta \\
\rho &= 3P + \frac{c^4}{8\pi G_N} G_\mu^\mu \\
u_0^2 &= \frac{(T_{00} - g_{00} P)}{(\rho + P)} = -\frac{\frac{c^4}{8\pi G_N} G_{00} + g_{00} P}{(\rho + P)} \\
u_r^2 &= \frac{(T_{rr} - g_{rr} P)}{(\rho + P)} = -\frac{\frac{c^4}{8\pi G_N} G_{rr} + g_{rr} P}{(\rho + P)}
\end{aligned}
\tag{23}
$$

This allows that for any given geometry expressed in terms of its metric and Einstein tensor, the physical parameters for the pressure $P$, density $\rho$, and four velocity $\boldsymbol{u}$ can be directly determined. When the black hole geometries previously discussed are inserted into these equations, the fluid parameters are tachyonic and/or do not generally satisfy the consistency condition normalizing the 4 velocities. However, one expects a cosmology describing the universe as a whole should consistently define such physical parameters. One is able to show that for a fluid dynamic cosmology, the density dynamics satisfies

$$
\frac{d\rho}{dct} = -\sqrt{\frac{24\pi G_N\, \rho}{c^4}} (\rho + P). \tag{24}
$$

The form of this equation is completely consistent with that described by big bang cosmology and the Friedman-Lemaitre equations of standard cosmology, without reference to a particular representation of the geometry. In the final section, a dual scaled geometry will be developed to describe the physical parameters without requiring the constancy of the cosmologically coherent term in the model.

## Dynamics of a Cosmological Constant

It is instructive at this point to examine the large scale structure of geometry describe by only a (positive) cosmological constant. This geometry can be presented in various forms, including the static de Sitter form and the Robertson-Walker form. The metrics for these two representations can be written

$$ds^2 = -\left(1-\frac{r_{dS}^2}{R_\Lambda^2}\right)(dct_{dS})^2 + \frac{dr_{dS}^2}{\left(1-\frac{r_{dS}^2}{R_\Lambda^2}\right)} + r_{dS}^2 d\omega^2 = -(dct_{RW})^2 + e^{2ct_{RW}/R_\Lambda}\left(dr_{RW}^2 + r_{RW}^2 d\omega^2\right) \tag{25}$$

Analytic expressions can be developed relating the two coordinate representations. The de Sitter temporal and radial coordinates can be expressed using those of the Robertson-Walker form,

$$\begin{aligned} ct_{dS} &= ct_{RW} - \frac{R_\Lambda}{2}\log\left(1-\frac{r_{RW}^2}{R_\Lambda^2}e^{2ct_{RW}/R_\Lambda}\right) \\ r_{dS} &= r_{RW}\, e^{ct_{RW}/R_\Lambda} \end{aligned} \tag{26}$$

while the inverse relationships satisfy

$$\begin{aligned} ct_{RW} &= ct_{dS} + \frac{R_\Lambda}{2}\log\left(1-\frac{r_{dS}^2}{R_\Lambda^2}\right) \\ r_{RW} &= \frac{r_{dS}\, e^{-ct_{dS}/R_\Lambda}}{\left(1-\frac{r_{dS}^2}{R_\Lambda^2}\right)^{1/2}} \end{aligned} \tag{27}$$

The Penrose diagrams using the two coordinate representations are demonstrated in FIGURE 12:

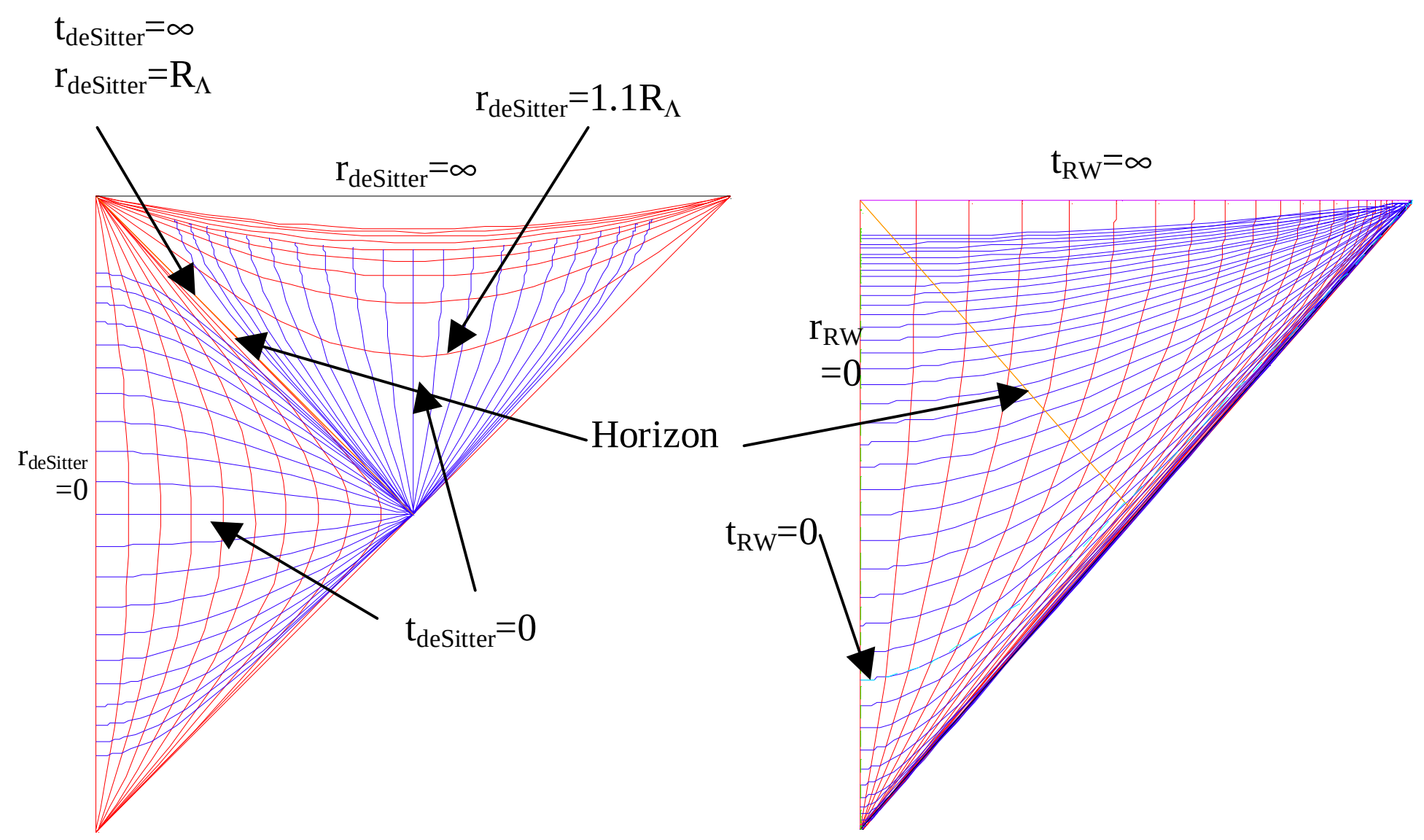

**FIGURE 12** Penrose diagrams for a space-time of a cosmological constant, expressed in de Sitter and Robertson-Walker coordinates.

To causally complete the space, the diagrams can be reflected about the diagonal connecting the lower left corner to the upper right corner, as was done in the Schwarzschild space-time Penrose diagram FIGURE 2. The horizon is static using the de Sitter coordinates, but not so using the Robertson-Walker coordinates. However, in both cases it is given by the ingoing light-like surface that reaches any given co-moving "center" of the geometry at infinite time. A Penrose diagram using coordinates analogous to those used in the right hand diagram in FIGURE 12 for the dynamic cosmology observed is a goal of the approach being described in this paper.

## Friedman-Robertson-Walker-Lemaitre Cosmology

A brief overview of some aspects of the standard model of cosmology will be presented in this section. The *cosmological principle* asserts that our position in the universe is no more special than any other, and assumes a spatially homogeneous and isotropic cosmology. Such a cosmology can be represented using the general coordinates of the Friedman-Robertson-Walker (FRW) metric form:

$$ds^2 = -c^2 dt^2 + R^2(t)\left(\frac{dr^2}{1-\kappa r^2} + r^2 d\vartheta^2 + r^2 \sin^2\vartheta d\varphi^2\right). \qquad (28)$$

By convention, in this form the FRW scale *R(t)* is taken to carry all dimension of length in the geometry, and the spatial curvature $\kappa$ is a dimensionless constant equal to +1 ,-1, or 0 for closed, open, or spatially flat, geometries.

Assuming that an ideal fluid drives Einstein's equation with a cosmological constant for this geometry, one obtains the Friedman-Lemaitre equations for standard cosmology:

$$\left(\frac{\dot{R}}{R}\right)^2 = \frac{8\pi G_N}{3c^2}\left(\rho + \rho_\Lambda\right) - \frac{\kappa c^2}{R^2}$$
$$\frac{\ddot{R}}{R} = -\frac{4\pi G_N}{3c^2}\left(\rho + 3P - 2\rho_\Lambda\right) \qquad (29)$$

where the cosmological constant is contained in $\rho_\Lambda \equiv \frac{\Lambda}{8\pi G_N}$. These equations are consistent with observed cosmology and adiabatic expansion during the big bang.

## A Dynamic Metric with Multiple Scales: Two-Fluid Cosmology

As previously mentioned, the observed accelerating expansion in causally disparate regions of the universe introduces a dimensional scale in the cosmology. The Friedman-Lemaitre Equations (29) are spatially scale invariant, exclusive of the cosmological constant. However, to describe general dynamics, one desires a metric form that incorporates cosmological scales with differing temporal dependencies. In this section, non-orthogonal coordinates will be developed so as to mix a scale that would become like de Sitter space when orthogonalized using the temporal coordinate with one that would become like Robertson-Walker space when orthogonalized using the radial coordinate. This dual-scaled cosmology should share dynamic aspects of both types of space-times, yet be consistent with the fluid cosmology that describes the big bang. The form of the mixed coordinate metric to be considered is given by[9]

$$ds^2 = -\left(1 - \frac{{r_{th}}^2 R_{th}^2(ct)}{R_C^2(ct)}\right)(dct)^2 - \frac{2 r_{th} R_{th}^2(ct)}{R_c(ct)} dct\, dr_{th} + R_{th}^2(ct)\left[{dr_{th}}^2 + {r_{th}}^2\left(d\theta^2 + \sin^2\theta\, d\varphi^2\right)\right]. \qquad (30)$$

This form is seen to be like a Robertson-Walker metric for the observer at $r_{th}=0$. Likewise, if the scale factors $R_{th}$ and $R_c$ are constant, the metric can be orthogonalized as was done for Eq. (8) to yield de Sitter space. Therefore, this metric form incorporates a scale $R_C$ that will approximate or reproduce a cosmological *constant* in Einstein's Equation. Such a form allows dynamic evolution of dark energy and potential quantum evolution of cosmology. It can be shown to be a very convenient form for studying early cosmology[15].

The two scales in Eq. (30) provide physical parameters for a two-fluid cosmology. One can define a coherent density scale $\rho_{coherent}$ in terms of the coherent cosmological scale $R_c$ in a manner analogous to the relationship between dark energy density and a cosmological constant $\rho_\Lambda \equiv \frac{\Lambda}{8\pi G_N} = \frac{3c^4}{8\pi G_N}\left(\frac{1}{R_\Lambda}\right)^2$. The total energy density in Eq. (23) then decomposed into two components according to

$$\rho = \rho_{thermal} + \rho_{coherent} \quad , \quad \rho_{coherent} = \frac{3c^4}{8\pi G_N}\left(\frac{1}{R_c}\right)^2 . \qquad (31)$$

Defining a combined scale $R$ that satisfies $\frac{\dot{R}}{R} = \frac{\dot{R}_{th}}{R_{th}} + \frac{1}{R_c}$, Einstein's equation for the geometry defined by Eq. (30) yields

$$\left(\frac{dR/dt}{R}\right)^2 = \left(\frac{dR_{th}/dt}{R_{th}} + \frac{c}{R_c}\right)^2 = \frac{8\pi G_N}{3c^2}\left(\rho_{thermal} + \rho_{coherent}\right). \qquad (32)$$

The thermal scale can be seen to evolve proportionate to the evolution of the combined scale, with a proportionality related to the difference of the coherent fraction of energy density from unity[10]:

$$\frac{dR_{th}/dt}{R_{th}} = \sqrt{\frac{8\pi G_N}{3c^2}}\left[\sqrt{\rho} - \sqrt{\rho_{coherent}}\right] = \left(\frac{dR/dt}{R}\right)\left[1 - \sqrt{\frac{\rho_{coherent}}{\rho}}\right]. \qquad (33)$$

Rather than choosing to orthogonalize the metric Eq. (30) by redefining a temporal parameter as was done for the black hole geometry examined earlier[3**Error! Bookmark not defined.**], coordinates will be developed that preserve the temporal parameter and redefine a radial parameter. The form that connects the two coordinate representations can be determined in a straightforward manner to satisfy[11]

$$\begin{aligned} &r\,R(ct) = r_{th}\,R_{th}(ct), \\ &ds^2 = -(dct)^2 + R^2(ct)\left[dr^2 + r^2\left(d\vartheta^2 + \sin^2\vartheta\, d\varphi^2\right)\right], \end{aligned} \qquad (34)$$

which provides a Robertson-Walker coordinate representation for a geometry with vanishing spatial curvature $\kappa=0$. The temporal coordinate is just the proper time coordinate of a co-moving observer in the cosmology. The geometry maintains its spatial scale invariance in terms of the combined cosmological scale $R$. There is no need to include a separate cosmological constant term in Einstein's equation, since a cosmologically coherent scale (which can be, but need not be, temporally constant) has been introduced via $R_c$ . Given the time dependence of the coherent fraction, it is straightforward to construct a Penrose diagram for the given space-time. In addition, the evolutionary histories and futures of the thermal and coherent scales can be determined once the equations of states relating the pressures to the energy densities of the respective fractions are specified. The decomposition presented should be useful in examining phase transitions in the early universe[12,13,14]. In addition, since transitions between coherent and thermal physical contents can be directly explored using this approach, one can examine the operational beginnings of time[15]. Therefore, it does seem that examinations of non-orthogonal temporal coordinates can provide some insights into the evolution of cosmologies of interest, especially in regions that would be singular using dynamics specified in terms of more obvious temporal coordinates.

## DISCUSSION AND CONCLUSIONS

The discussion presented here has focused on the use of temporal coordinates that are not orthogonal to the radial coordinates near the horizons that manifest in various geometries. Although the spatio-temporal coordinates used *are* orthogonal for important classes of observers, like distant observers of black holes or local observers on co-moving centers in dynamic cosmologies, the non-orthogonal behaviors do not introduce singular physical content due to anomalous coordinate behaviors. One finds that a naïve introduction of temporal dependencies guided from static geometries into the physical parameterization of dynamic horizons introduces physically singular behavior at the (previously static) horizons. In such geometries, any observer would describe some sort of *brick wall* at the horizons. This is somewhat discomforting, since *any* in-falling observer to a very very large black hole must cross and experience a surface with singular content prior to continuing toward the singular center of the black hole through space-time of relatively uninteresting and mild curvature. Complementarity[1] asserts that no observer witnesses a violation of any law of nature. In particular, an inertial observer should not detect the presence of any accelerating observer's coordinate anomalies. Clearly, the temporal dependency of a horizon scale has been seen to qualitatively modify the local coordinate structure of the space-time. Therefore, coordinates have been chosen that parameterize dynamics in a way that is non-singular near the horizons.

In addition, the "Asymptopias" of dynamic black holes seem to differ from those of static black holes. While both the static Schwarzschild black hole and the dynamic black holes discuss in this paper have geometries that are asymptotically flat Minkowski space-times, they are different Minkowski space-times. The horizon of a static Schwarzschild geometry is a $t=\infty$ surface, while those of the dynamic black holes have surfaces of constant $t$ crossing the dynamic horizon. The past and future light-like infinities $\Im^+$ and $\Im^-$ in the dynamic geometries that correspond with Minkowski space-times cannot be the same surfaces as those in any rescaled static Schwarzschild geometry while maintaining the horizon as an asymptotic time surface. Mathematically, the asymptotic behavior of the dynamic conformal coordinates has been found to be qualitatively different for a vanishing value of the rate of mass accretion/evaporation.

A preliminary Penrose diagram for the complete life cycle of a black hole has been presented. Generally, the horizon of a dynamic black hole need not coincide with a coordinate singularity. This is the primary reason that physical singularities are not needed to generate non-existent geometric singularities in the dynamic geometries discussed. The dynamics and locations of the radial mass scale relative to the horizon of the black hole during accretion and evaporation epochs was as expected. However, the center $r=0$ of the geometry undergoes a light-like transition as the singularity forms in a manner whose diagrammatic representation was unexpected. In hindsight, this transition was necessary in order to expand the space-time in a manner that could self-consistently mesh the nearby black hole geometry with that of the asymptotic Minkowski space-time of a distant observer.

The behavior of a quantum field in the background space-time of an evaporating black hole was also discussed. Quantum coherence has been experimentally verified in near earth gravitating systems[16]. For the system presented, it was assumed that the quantum field had negligible affect upon the background geometry. Since the geometry itself behaves in a spatially coherent manner, it should be possible to determine a quantum field that satisfies the appropriate boundary conditions and generates the geometry. The behavior of quantum fields in the geometry is being actively examined by the author, and will be submitted in a future publication.

Finally, another metric with non-orthogonal local spatio-temporal coordinates was presented to incorporate gravitational coherence on a cosmological scale. The metric was seen to be transformable into a Robertson-Walker form using that same temporal parameter. Yet, the coherent scale introduced allows for the incorporation of static or evolving "dark energies" while maintaining the overall spatial scale invariance of the cosmology. Investigations have begun towards developing a Penrose diagram for a cosmology satisfying the dynamic equations with coherent and thermal energy scales.

## ACKNOWLEDGMENTS

The author gratefully acknowledges useful discussions with Stephon Alexander, James Bjorken, Beth Brown, Tihani Finch, E.D. Jones, Harry Morrison, H. Pierre Noyes, Michael Peskin, Paul Sheldon, and Lenny Susskind.